\pdfoutput=1 
\documentclass[notoc]{JINST}

\usepackage{xspace}
\usepackage{url}
\usepackage{paralist}
\newcommand{\apenetp}{APEnet+\xspace}
\newcommand{\apelink}{APElink\xspace}
\newcommand{\apenet}{APEnet\xspace}

\newcommand{\ie}{\textit{i.e.}\xspace}
\newcommand{\eg}{\textit{e.g.}\xspace}
\newcommand{\pcie}{PCIe\xspace}
\newcommand{\apenetv}{\mbox{APEnet v5}\xspace}
\newcommand{\nios}{\texttt{Nios~II}\xspace}

\newcommand{\muC}{$\mu$C\xspace}

\title{Architectural Improvements and Technological Enhancements for the
\apenetp Interconnect System.}

 \author{R. Ammendola$^a$, 
         A. Biagioni$^b$\thanks{Corresponding author.}~, 
		 O. Frezza$^b$, 
		 A. Lonardo$^b$, 
		 F. Lo Cicero$^b$, 
		 M. Martinelli$^b$, 
         P.~S.~Paolucci$^b$, 
		 E. Pastorelli$^b$, 
		 D. Rossetti$^c$, 
		 F. Simula$^b$, 
		 L. Tosoratto$^b$, 
		 P. Vicini$^b$\\
\llap{$^a$}INFN Sezione di Roma Tor Vergata,\\
  Via della Ricerca Scientifica, 1 - 00133 Roma Italy\\
\llap{$^b$}INFN Sezione di Roma\\
  P.le Aldo Moro, 2 - 00185 Roma Italy\\
\llap{$^c$}NVIDIA Corp\\
  2701 San Tomas expressway\\
  Santa Clara, CA 95050\\\  
  
  E-mail: \email{andrea.biagioni@roma1.infn.it}}

\abstract{The \apenetp board delivers a
\mbox{point-to-point}, \mbox{low-latency}, 3D torus
network interface card. In this paper we describe the latest
generation of \apenet NIC, \apenetv, integrated in a \pcie Gen3 board
based on a \mbox{state-of-the-art}, 28~nm Altera Stratix V FPGA. The
NIC features a network architecture designed following the Remote DMA
paradigm and tailored to tightly bind the computing power of modern
GPUs to the communication fabric. For the \apenetv board we show
characterizing figures as achieved bandwidth and BER obtained by
exploiting new high performance ALTERA transceivers and \pcie Gen3
compliancy.}


\begin{document}

\section{Introduction}
\label{sec:intro}

The \apenetp project delivered a \mbox{point-to-point} \pcie Gen2
interconnect adapter to be employed in hybrid x86+GPU computing
clusters with a 3D toroidal network mesh.
\apenetp high performance and low latency capabilites are the result
of a network architecture designed following the Remote DMA paradigm
which is tailored to tightly bind the computing power of modern GPUs
to the communication fabric.
Doing so on the \apenetp board was possible by employing latest
standards for the physical interconnect --- \mbox{SFF-8436} with its
QSFP+ modules --- and exploit them by means of the large hardware
resources provided by a \mbox{state-of-the-art} FPGA platform like the
Stratix IV.

Following up with enhancements in these areas --- the most significant
being an upgraded interconnect standard (\mbox{SFF-8665}) with new
zQSFP plugs and the latest generation for the \pcie standard --- means
employing an evolved FPGA platform able to integrate and exploit these
improvements alongside added functionality.
This requires a redesign of the NIC to support the new
\pcie Gen3 protocol, faster transceivers and a much larger amount of
programmable resources.
The chosen platform for this redesign exploration is the Stratix V
FPGA Development Kit.
The \apenetv is based on a 28~nm technology FPGA
(\mbox{5SGXEA7K2F40C2N}) and offers a \pcie $\times$8 connector, two
HSMC connectors for exporting high bandwidth links over a
daughterboard, a QSFP connector with optical cage and an Ethernet PHY
\mbox{10/100/1000Mbps} copper connector.
A \mbox{PLDA-proprietary} component --- QuickPCIe Expert --- was chosen as
a \mbox{Gen3-compliant} core interface towards the \pcie bus.
With this core, either an Avalon or an AXI bus can be selected as user
interface;
the choice of implementing the latter gives us room for preparing the
\apenetv environment to deal with the impact brought by a
\mbox{full-featured} ARM processor which is slated to be one of the
most significant additions on FPGA platforms of the next generation.
The new PLDA core also allows for greater freedom when choosing among
different strategies for the DMA channels management.
%

%
In section~\ref{sec:apenetplus} we summarize the main feature of the
\apenetp: in section~\ref{sec:hw} the very latest developments on the
\apenetv board on a 28~nm technology FPGA are reported --- harnessing the
complexities of a \pcie Gen3 interface being a so \mbox{far-reaching}
work in a GNU/Linux device driver as to deserve a specific and 
\mbox{in-depth} examination in the dedicated section~\ref{sec:sw}. 
In conclusion in section~\ref{sec:test} preliminary results obtained
with the \apenetv board are shown.

\section{\apenetp and related work}
\label{sec:apenetplus}

\apenetp~\cite{APEnetChep:2012} is a \mbox{point-to-point}, 
\mbox{low-latency} network controller developed by INFN for a 
\mbox{3D-torus} topology integrated in a \pcie Gen2 board based on an
Altera Stratix IV FPGA.
It is the building block for the QUonG~\cite{ammendola2011quong:long}
hybrid \mbox{CPU/GPU} HPC cluster and the basis for a
\mbox{GPU-enabling} data acquisition interface in the \mbox{low-level}
trigger of NA62, a High Energy Physics experiment at
CERN~\cite{NaNetTwepp:2013}.
The board provides 6 QSFP+ modules which are directly connected to the
Altera FPGA embedded transceivers.
Each transceiver is capable of a data rate up to 8.5~Gbps in fully
bidirectional mode; a single remote data link --- the
\apelink~\cite{APEnetTwepp:2013} --- is built up by bonding 4
transceivers composing a link operating at up to 34~Gbps.
The NIC is also able to directly access the memory of Fermi- and
\mbox{Kepler-class} NVIDIA GPUs implementing GPUDirect V2
(\mbox{peer-to-peer}) and the more recent GPUDirect RDMA
capabilities~\cite{ammendola:2013:GPU}.

In the field of reconfigurable computing, a fast communication path
between GPUs and FPGA-based devices is desirable.
In \cite{bittner:2012:Direct} a simpler implementation for connecting
GPU and FPGA devices directly via the \pcie bus is described, enabling
the transfer of data between these heterogeneous computing units
without the intermediate use of system memory.
In \cite{Hanawa:2013:TCA:2510648.2510972} a \mbox{GPU-to-GPU} communication mechanism via
FPGA is described with a different approach to address mapping which
does not require specific on-board memory management logic.
In \cite{Thoma:FPGA2} is presented an open source framework enabling
easy integration of GPU and FPGA resources providing direct data
transfer between the two platforms with minimal CPU coordination at
high data rate and low latency.

A recent development for major FPGA vendors and board integrators is
to come out with development kits and products with updated \pcie Gen3
host interfaces and multiple 10G/40G \mbox{off-board} links.
In \cite{NetFPGAsume} a \pcie \mbox{Gen3-compliant} NIC with the
stated goal of supporting 100~Gbps speeds is presented.
However, a lot of work concerning the software device driver and
hardware logic is required to exploit the enormous throughput assured
by these standards.
Research in data acquisition experiments, FPGA accelerators for
reconfigurable computing and high speed network interconnect for HPC
would benefit from this evolution.


\section{\apenetv architecture on new generation 28~nm FPGA}
\label{sec:hw}

\begin{figure}[!hbt]
  \begin{minipage}[t]{.5\textwidth}
    \centering
    \includegraphics[trim=30mm 5mm 30mm 20mm,clip,width=\textwidth]{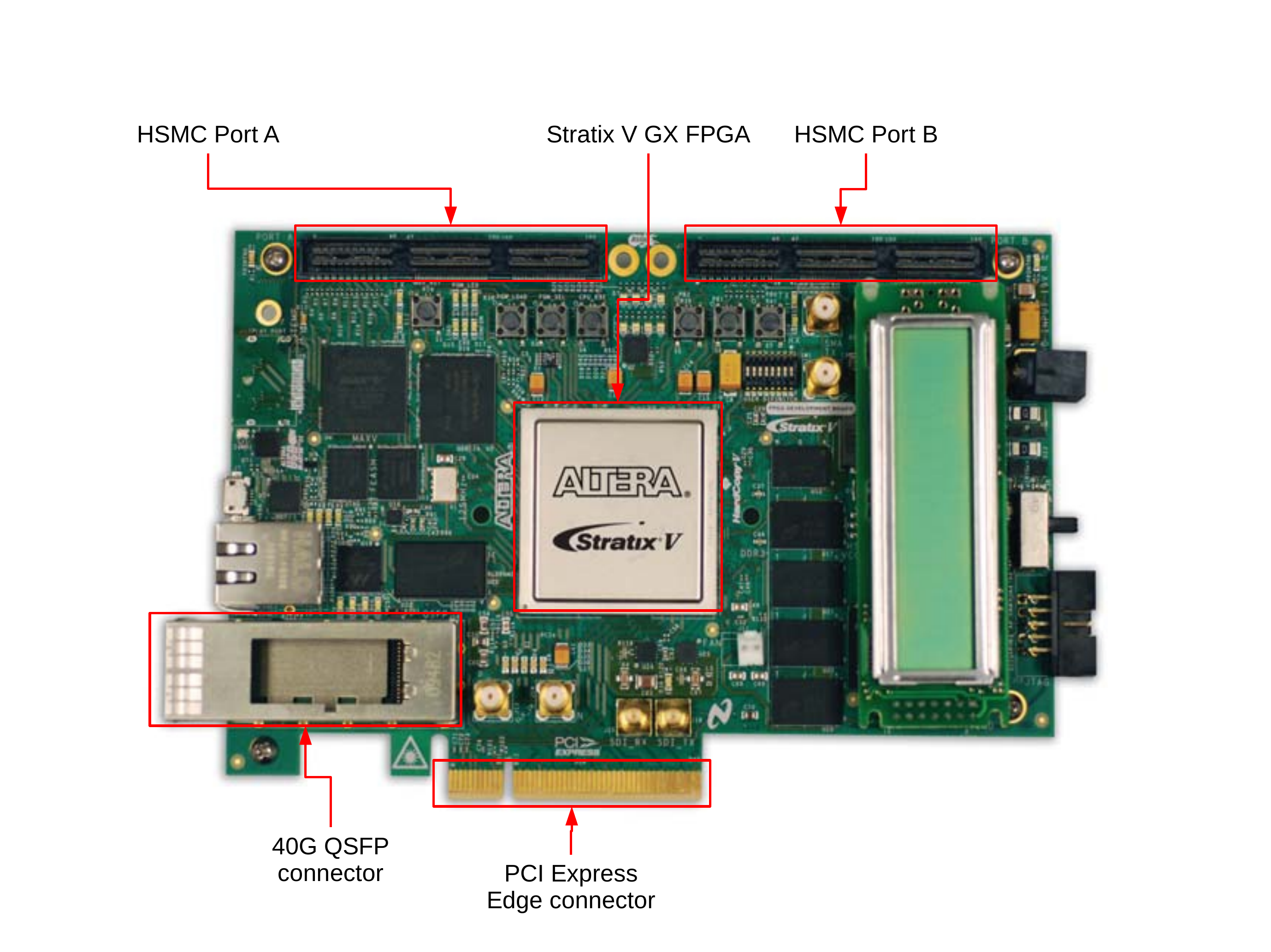} 
    \caption{Altera 28~nm Stratix V GX Development Board.}  
    \label{fig:SV_board}\end{minipage}
  \quad
  \begin{minipage}[t]{.5\textwidth}
    \includegraphics[trim=60mm 20mm 60mm 20mm,clip,width=.85\textwidth]{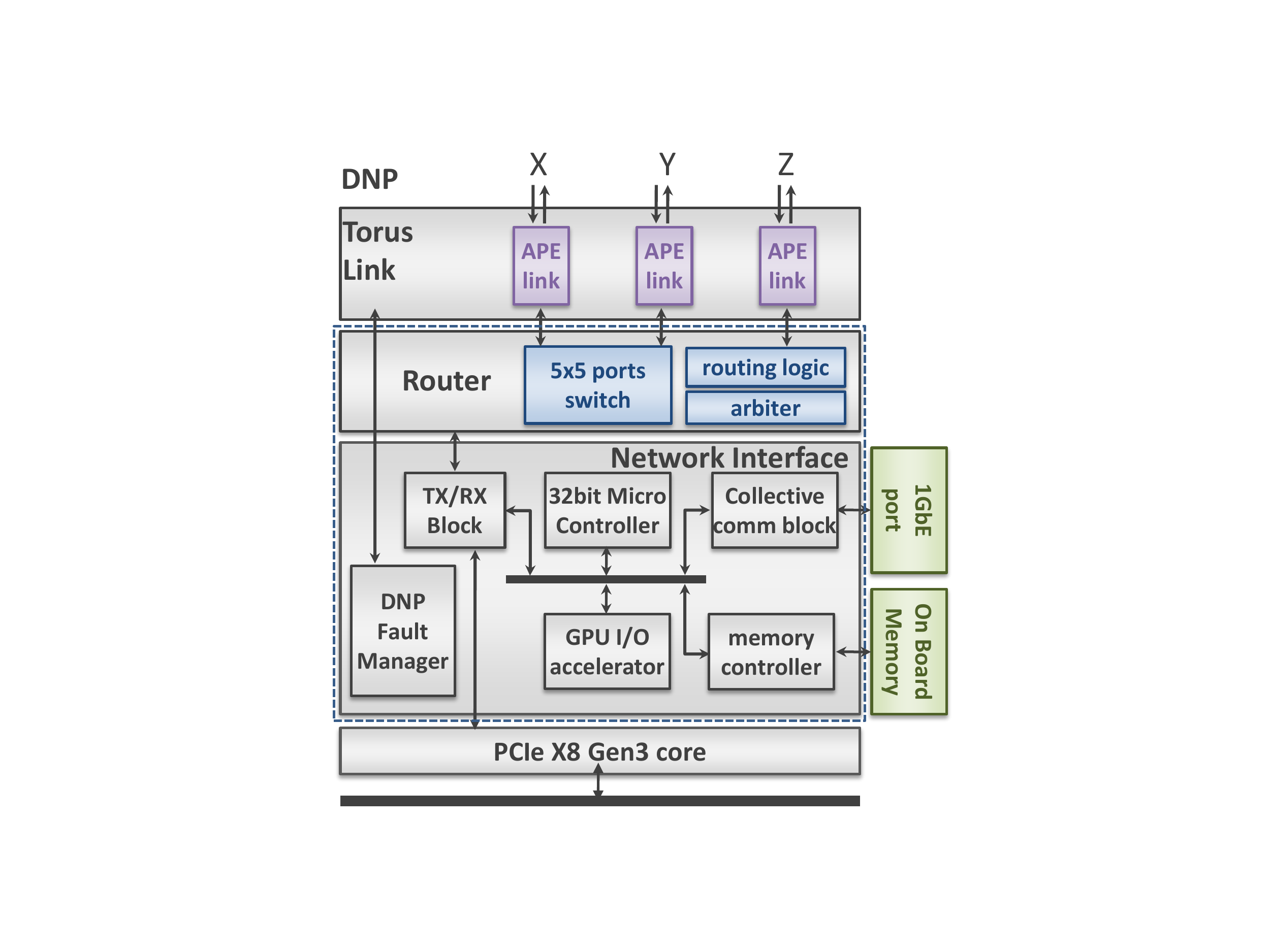}
    \caption{\apenet architecture on Stratix V}
    \label{fig:apenet_arch_V5}
  \end{minipage}
\end{figure}

\apenetv is the name of the latest generation \apenet board based on
the Altera \mbox{DK-DEV-5SGXEA7N} development kit
(figure~\ref{fig:SV_board}), a complete design environment featuring a
28~nm FPGA.
Two major improvements stem from adoption of the Stratix V: \pcie Gen3
interface and new embedded transceivers with a data rate up to
14.1~Gbps for increased \mbox{off-board} link speed.

The Stratix V FPGA offers a \pcie Gen3 connector supporting a data
rate of \mbox{8.0~Gbps/lane}.
The improved \mbox{128B130B} encoding scheme guarantees a host
interface bandwidth enhancement.
The total raw bandwidth that can be achieved with a $\times$8
interface is $\sim$ \mbox{7.9~GB/s}, to compare with the integrated
theoretical peak bandwidth of \mbox{4.0~GB/s} allowed by the Stratix
IV on board \apenetp.
As regards \mbox{off-board} connectivity, the Altera board sports a
40G QSFP connector and two HSMC ports.
The QSFP connector exploits 4 embedded Altera transceivers from the
FPGA device while the HSMC connectors --- called port A and port B ---
provide 8 lanes in port A and 4 lanes in port B with limited data rate
transceivers (10.0~Gbps).

The Distributed Network Processor (DNP) is the core of the \apenetv
architecture~\ref{fig:apenet_arch_V5}.
The DNP acts as an offloading engine for the computing node,
performing \mbox{inter-node} data transfers.
The Torus Link block manages the data flow by encapsulating packets
into a light, \mbox{low-level} word stuffing protocol able to detect
transmission errors via CRC.
It allows for a \mbox{point-to-point}, \mbox{full-duplex} connection
of each node with three neighbours.
The Router component is responsible for data routing and dispatching,
dynamically interconnecting the ports of the \mbox{cross-bar} switch;
it is able to simultaneously handle 5 flows @5.6 GB/s applying a
\mbox{dimensioned-ordered} routing policy with Virtual
\mbox{Cut-Through}~\cite{Kermani79virtualcut-through:} algorithm.
Finally, the Network Interface is the packet injection/processing
logic; it manages data flow to and from either Host or GPU memory.
On the receive side, it provides hardware support for the Remote
Direct Memory Access (RDMA) protocol~\cite{Ammendola:2013:FPT},
allowing remote data transfer over the network without involvement of
the CPU of the remote node.
An integrated \muC provided by the FPGA allows for
straightforward implementation of RDMA semantics.

\subsection{PCI Express interface}

\begin{figure}[!hbt]
  \begin{minipage}[t]{.47\textwidth}
    \centering
    \includegraphics[trim=0mm 10mm 0mm 5mm,clip,width=\textwidth]{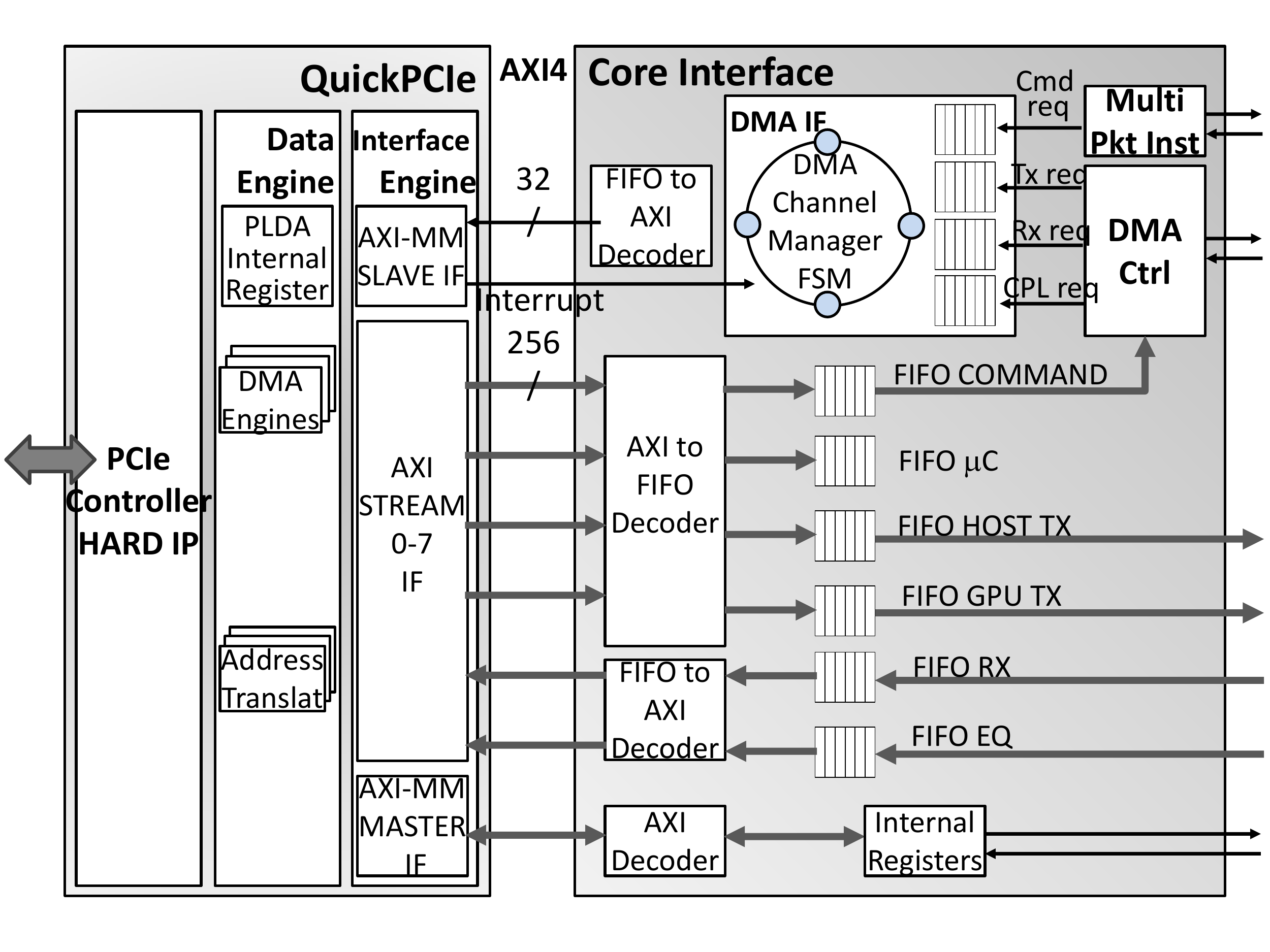}
    \caption{Architecture of APEnet Gen3 on Stratix V FPGA.}
    \label{fig:PCIe_V5}
  \end{minipage}
  \quad
  \begin{minipage}[t]{.47\textwidth}
    \includegraphics[trim=0mm 8mm 0mm 8mm,clip,width=\textwidth]{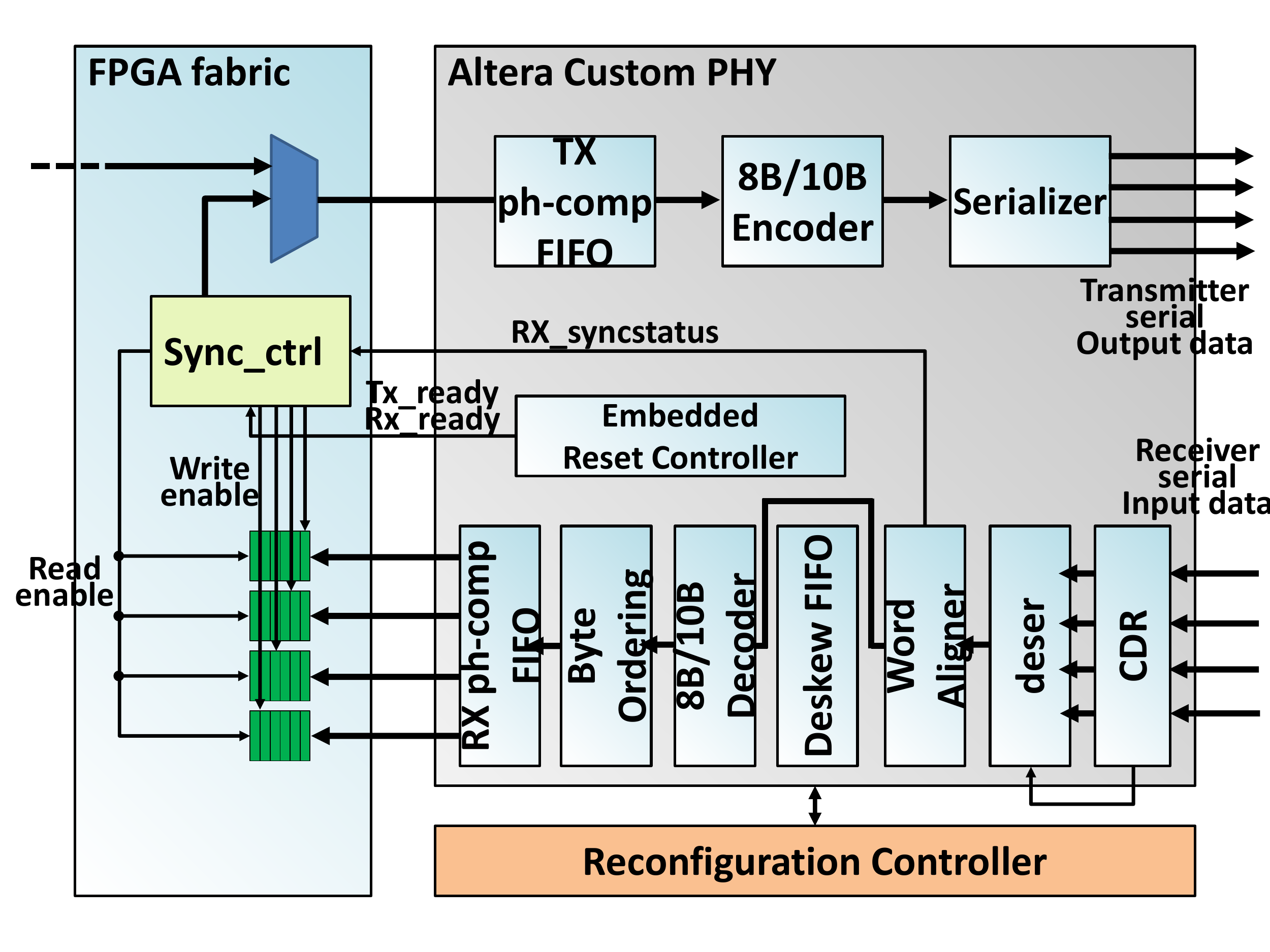}
    \caption{\apenetv Torus Link block architecture.}
    \label{fig:stratixV_link}
  \end{minipage}
\end{figure}

A redesign of the \pcie interface is mandatory to exploit the Gen3
capabilities provided by the latest Altera FPGA generation.
The main components of the interface are the \textit{QuickPCIe Expert
IP}, a \mbox{PLDA-proprietary} component, and an overhauled version of
the \mbox{TX/RX} block (figure~\ref{fig:PCIe_V5}).

The PLDA core wraps around Altera PCI Express Hard IP allowing for
access to the \pcie configuration space.
It interconnects and arbitrates between input and output flows
implementing up to 8 independent DMA Engine modules and converting
between the \pcie and AMBA AXI interfaces.
Current implementation of the \apenetv provides an \mbox{AXI4-Lite}
Slave Interface used to access PLDA internal registers, an
\mbox{AXI4-Lite} Master Interface for \apenetp registers, four AXI4
Stream Input Interfaces and two AXI4 Stream Output Interfaces.
The AXI interface will simplify the connection of the ARM processor,
which could replace our work with the proprietary Altera \nios
microcontroller on future revisions of the device.
The AXI4 Memory Mapped slave interface is connected to the DMA\_IF
inside the \mbox{TX/RX logic}.
It programs direct DMA transfers implementing four fully independent
DMA Engine Modules in order to take commands used to instantiate
memory read transactions, to access the \mbox{Host/GPU} memory and to
communicate event completions.
It sequentially pops DMA requests and programs DMA transfers writing
in PLDA Configuration Space Registers.
It is made aware of when transactions are complete by an interrupt
which is issued on the AXI domain by the QuickPCIe DMA engines.
AXI4 Stream interfaces are used to push/pop data, command and
completion word in and out to FIFOs.


%

\subsection{\mbox{Off-board} interface} 

Stratix V GX FPGAs feature \mbox{full-duplex} transceivers with data
rates from 600~Mbps to 14.1~Gbps, offering several programmable and
adaptive equalization features.

The Physical Layer of Torus Link block is made up of an Altera Custom
IP Core (with the corresponding reconfiguration block) and a
proprietary channel control logic (Sync\_ctrl block), as shown in
figure~\ref{fig:stratixV_link}.
The Altera Custom IP Core is a generic PHY which can be customized in
order to meet design requirements.
The receiving side consists of a Word Aligner, 8B/10B decoder, Byte
Ordering Block and RX phase compensation FIFO.
Similarly, on the transmitter side, each transceiver includes TX phase
compensation FIFO, 8B/10B encoder and Serializer.
The Word Aligner restores the word boundary based on an alignment
pattern that must be received during link synchronization.
A status register asserted by the Altera Avalon interface triggers the
Word Aligner to look for the word alignment pattern in the received
data stream.
The Byte Ordering block looks for the byte ordering pattern in the
parallel data: if it finds the byte ordering pattern in the MSB
position of the data it inserts pad bytes to push the byte ordering
pattern to the LSByte(s) position.
Finally, the RX phase compensation FIFO compensates for the phase
difference between the parallel receiver clock and the FPGA fabric
clock.
The write and read enable signal of the Deskew FIFOs are managed by
the \texttt{Sync\_ctrl} block: the write signals are asserted for each
lane after recognition of 8B/10B keyword /K28.3/, while the read
signal, common to all FIFOs, is asserted when all FIFOs are no longer
empty.
Altera Transceiver Reconfiguration Controller dynamically reconfigures
analog settings in Stratix V devices: it is able to compensate for
variations due to process, voltage and temperature in 28~nm devices;

\begin{table}[hbt]
  \smallskip
  \centering
  \begin{tabular}{|c|cccc|}
    \hline
    \multicolumn{3}{|c|}{\textbf{Cable}}         & \textbf{BER}     & \textbf{Data Rate}       \\
    \hline
    \multicolumn{3}{|l|}{10~m Mellanox optical cable}  &  $<~$ 2.36 E-14 &  11.3 Gbps               \\
    \multicolumn{3}{|l|}{ 1~m Mellanox copper cable}   &  $<~$ 1.10 E-13  & 10.0 Gbps               \\
    \hline
  \end{tabular}
  \caption{\apenetp BER measurements on Altera 28~nm FPGA.}
  \label{tab:BERMeasure}
\end{table}

In conclusion, we implemented three \mbox{bi-directional} data
channels.
The $X$ channel was implemented using the 4 lanes of the
\mbox{40G-QSFP} connector; the $Y$ and $Z$ channels were implemented
onto the HSMC interface.
Very preliminary BER measurements were performed on the $X$ channel
of \apenetv (see table~\ref{tab:BERMeasure}).
The testbed consists of two Stratix V FPGAs connected by InfiniBand
cables with different lengths and support media (optical and copper).
Copper wires results derive from experiments conducted with
\mbox{10~Gbps-certified} cables; they are expected to improve as soon
as they are repeated with commercially available
\mbox{14~Gbps-certified} ones.
Moreover, the results are satisfactory, taking into account that no
analog parameters fine tuning was applied.

\section{\pcie Gen3 Driver Design}
\label{sec:sw}

All \apenetv software is developed and tested on the GNU/Linux x86\_64
platform and is available under the GNU GPL Licence.
A \mbox{low-level}, custom RDMA API is available, in principle similar
to, though much simpler than, Infiniband \textit{verbs} and the former
Myrinet \textit{GM}.
The RDMA APIs are available as a C language library, containing a
small set of functions and data types:

\begin{compactitem}
\item Communication primitives available to applications are:
  \texttt{rdma\_put/send()}.
 They are asynchronous --- simply pushing a command into a queue and
 returning when data have been read by the NIC --- and optionally
 \mbox{non-blocking}, \ie they return on the event of a full command
 queue, requiring to be reissued.
\item Memory buffer registration is mandatory for receive buffers and
  allows memory areas to be reserved for remote node access and to add
  their addresses (virtual and physical) to the NIC internal buffer
  table: \texttt{register\_buffer(), unregister\_buffer()}.
\item Events are generated on completion of communication primitives,
  when \mbox{pre-registered} receive buffers are written or when error
  conditions arise.
 The \texttt{wait\_event()} API call is used to wait for the arrival
 of an event with an optional \mbox{time-out}.
\end{compactitem}
For both communication and buffer registration APIs, memory buffers
are referenced by their virtual address --- \eg \texttt{rdma\_put()}
call takes as argument the virtual address of the remote memory
buffer.
As it is usual for the RDMA paradigm, there is no receive primitive as
receive buffers are always posted before their effective use --- \ie
posting is implicitly done by buffer registration.
Buffer registration, which is mandatory and explicit for receive
buffers, involves \mbox{memory-pinning} and \mbox{virtual-to-physical}
address translation and persists until an explicit buffer
deregistration.
Pinning and translation are implicitly done for the communication
primitives \texttt{rdma\_put/send()} as well as for the
\texttt{register\_buffer()} and \texttt{unregister\_buffer()}.

\subsection {Memory Read (TX) Process}
\label{sec:swtx}

On the TX side the memory read process requires the NIC to be
instructed by the driver by passing a DMA descriptor.
The descriptor contains all the relevant information to
program \apenetv data transfer --- \ie where the source data buffer is
located and its length, where it must be sent and the instructions to
generate the ''sent event'' completion. The device driver inserts the
descriptors in a circular list called \texttt{tx ring}.
The \mbox{DMA-capable} memory region containing the \texttt{tx ring}
is allocated by the device driver during the initializing phase and the memory 
start address is sent to the board by updating the \apenetv registers.

A mechanism based on updating a tx\_ring\_read and a
tx\_ring\_write pointer, ensures avoiding the memory overwriting
by the device driver and inconsistent data reading by the
hardware.

For efficiency reasons, the driver is able to store multiple
descriptors before updating tx\_ring\_write pointer accordingly.
In this way, the board will program a single DMA for all the
descriptors.

Since virtual memory maps a \mbox{non-contiguous} memory region in a
contiguous virtual address space, the buffer could be fragmented all
over physical memory.
Every transfer larger than a single page --- typically 4~KB --- is
managed by a \mbox{scatter-gather} (SG) list, where every element of
the list is a page of the buffer.
In this case, each page will be mapped in one tx ring descriptor, \ie
every descriptor has the associated physical address of one page of
the buffer.

Data of completion events are maintained in a cyclic queue called
\textbf{event queue} that is the same for both the ''sent'' event and
the ''received'' event.
Therefore, after updating the tx\_ring\_write pointer thus having
issued a data transfer, the driver starts polling the \textbf{event
queue} to wait for completion events.
Whenever new completion events are found, they are copied into a
software queue, ready to be popped by the application process that
invokes the ''wait\_event'' library function.

\subsection {Memory Write (RX) Process}
\label{sec:swrx}

%
%
The NIC on the receiving side \mbox{DMA-writes} data packets coming
from torus links into host memory.
There is no need of intermediate copies from kernel space to user
space or driver queries for buffer memory address.
This reduces the bus occupancy and the number of switches between
kernel space and user space while handling network traffic.

The buffer virtual address is used by the NIC as key to retrieve from
the internal table all needed buffer information (primarily the ID of
the owner process and the buffer physical address).
The first task of the receiving logic is thus scanning the buffers
pool registered by a running application to determine whether the
destination virtual address actually belongs to any of them \--- this
process is called Buffer Search (BSRC).

If a buffer is found, the pages that compose it are retrieved, their
physical addresses are resolved \--- \mbox{virtual-to-physical} (V2P)
translation \--- and a number of DMA
transactions are finally issued targeting these addresses.
In \apenetv the address translation mechanism is implemented in two
ways: in software, as a firmware on the assisting microcontroller, and
in hardware, by means of a Translation Lookaside Buffer (TLB)
accelerating the BSRC and V2P tasks.

Therefore, the association between virtual and physical address must
be registered beforehand to enable the address translation process.
%
%
In order to accomplish this task, the hardware exposes a set of registers
where the driver writes ''commands'' to register pages and buffers.
%

%
Data arrival is signaled by generating a ``received'' event,
which is managed in the same way of ``sent'' events as described
before.

\section{Preliminary Results}
\label{sec:test}

The testbed consisted of a X9DRG-VF SuperMicro server, hosting a
Stratix V FPGA development board, and a Tektronix TLA-7012 Gen3 Logic
Analyzer.

\begin{figure}[!hbt]
  \begin{minipage}[t]{.5\textwidth}
    \centering
     \includegraphics[trim=5mm 20mm 5mm 15mm,clip,width=\textwidth]{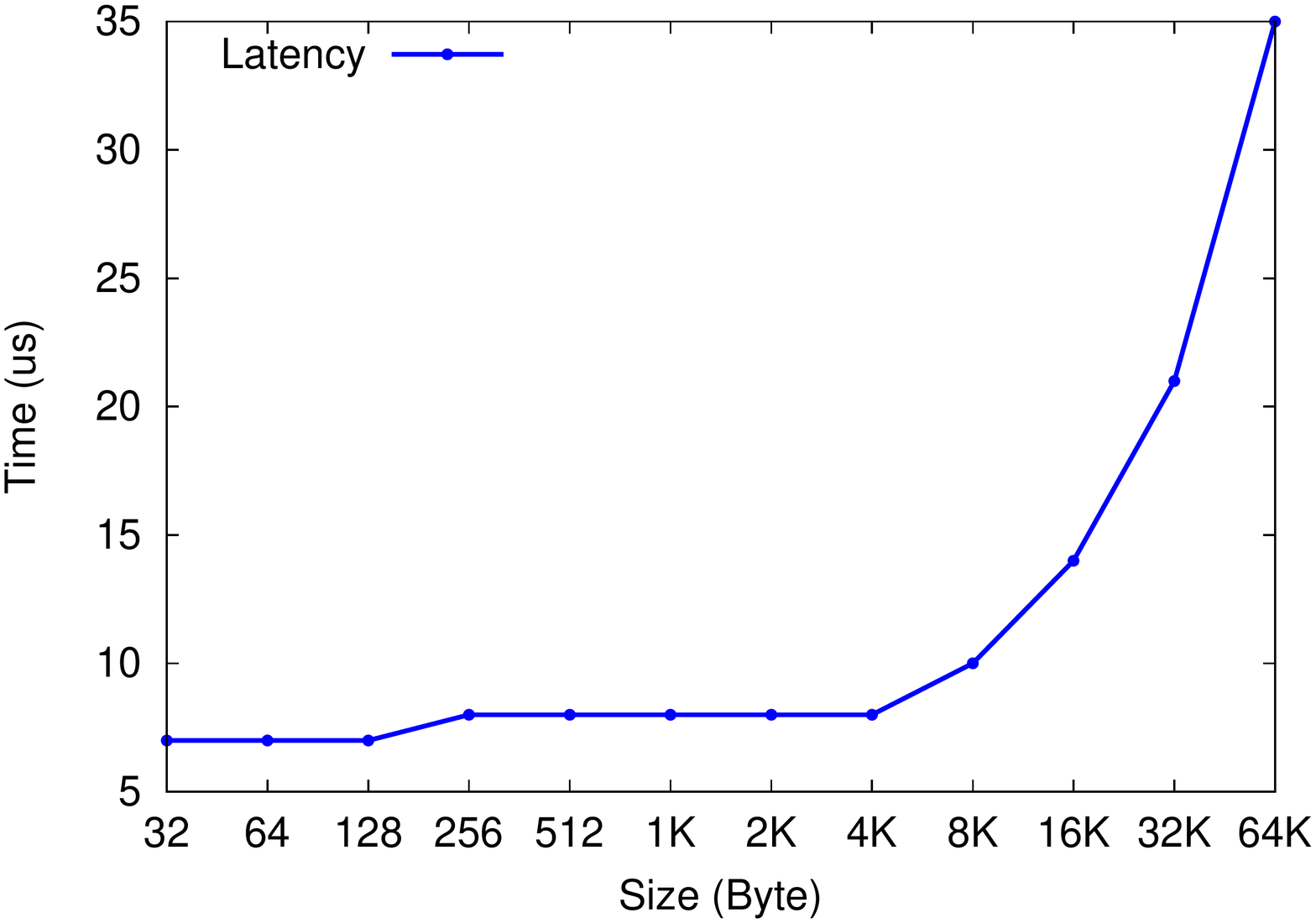}
    \caption{Latency loopback test.}
    \label{fig:gen3Latency}
  \end{minipage}
  \quad
  \begin{minipage}[t]{.5\textwidth}
     \includegraphics[trim=5mm 20mm 5mm 15mm,clip,width=\textwidth]{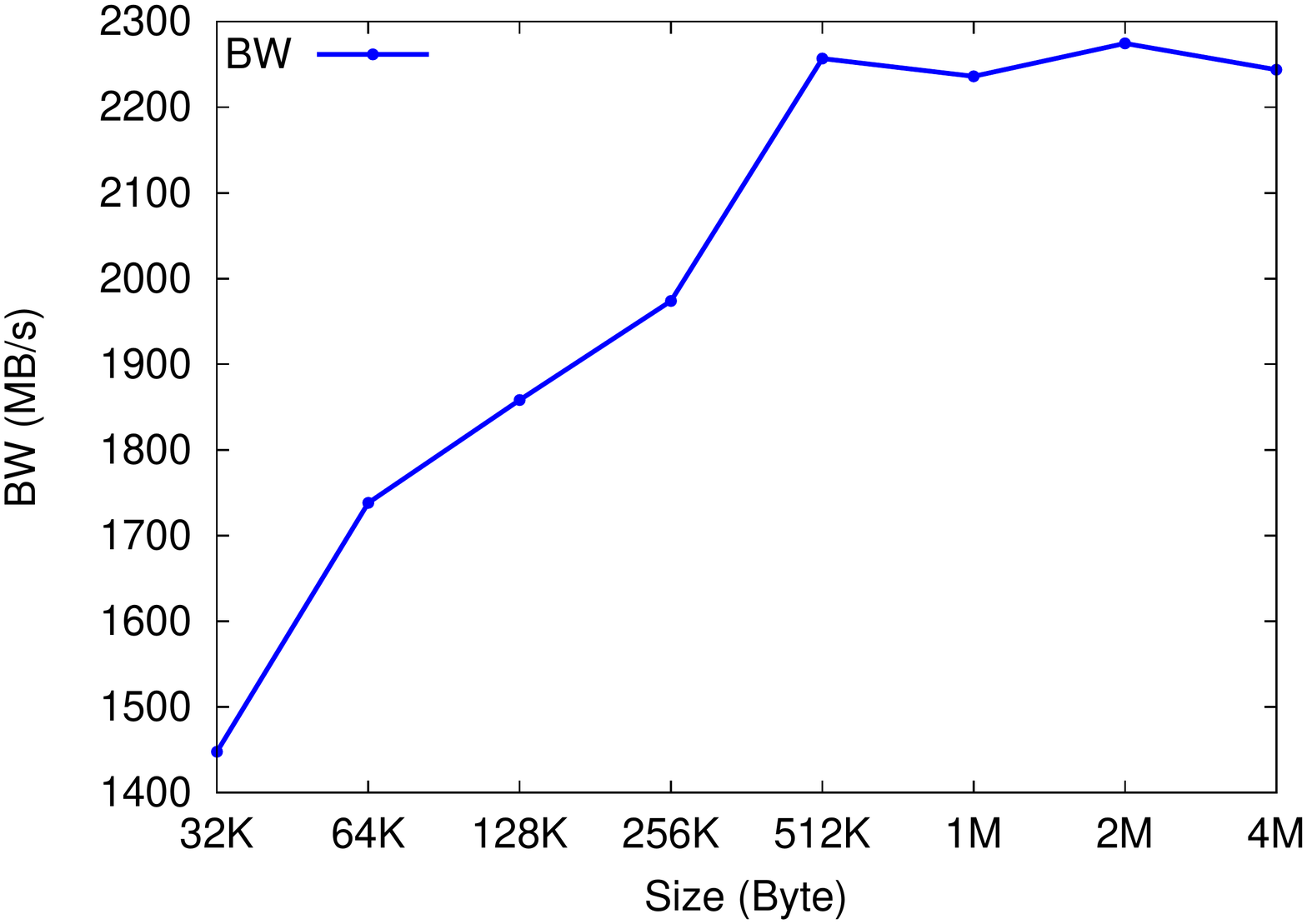} 
    \caption{Bandwidth loopback test.}
    \label{fig:gen3BW}
  \end{minipage}
\end{figure}

Two simple programs test the performance of the new architecture:
latency (Fig.~\ref{fig:gen3Latency}) is measured by sending a packet
of fixed size in \mbox{local-loop} and measuring the time required for
the ``sent'' event and the ``received'' event to arrive; the test is
repeated $\sim$100000 times and the resulting times are averaged.
In order to measure the bandwidth (Fig.~\ref{fig:gen3BW}) the test is
slightly different from the previous one: we send all the packets in a
once and then wait for all the sent and receive events.

Given that the development is still ongoing, latency results being
comparable with those achieved with the previous version of \apenetp
is already promising.
As regards the achieved bandwidth of 2.3~GB/s, we must observe that
the straightforward porting of \apenetp architecture onto Gen3 does
not regress from Gen2 but does not immediately yield significant
improvements either.
Indeed, a test performed using a reference design exploiting a single
DMA engine of the QuickPCIe Expert achieve 5.2~GB/s that represents
a target bandwidth for the \apenetv board. 

In table~\ref{tab:resources} are reported the resource consumptions
for the \apenetv firmware synthesized in an Altera Stratix V FPGA.  Current
design occupies a small fraction of the FPGA, leaving space for more
hardware optimizations. For example increasing the size of the TLB can
result in better figures for the RX process latency.

\begin{table}[hbt]
  \caption{\apenetv resource consumption on an Altera 28~nm FPGA.}
  \label{tab:resources}
  \smallskip
  \centering
  \begin{tabular}{|c|cccc|}
    \hline
    \textbf{Project}  & \textbf{Board}  & \textbf{ALMs} & \textbf{Register} & \textbf{Memory [MB]}  \\
    \hline
    \apenetv        & 5SGXEA7K2F40C2N & 76747 (33\%)              & 91447      &   1.55 (24\%)         \\
    \hline
  \end{tabular}
\end{table}

\section{Conclusions}
\label{sec:end}
We have presented the work done to advance the tested IP of an
\mbox{FPGA-based}, 3D toroidal network card like \apenetp to a 28~nm,
\mbox{Gen3-compliant} FPGA with its initial performance figures.

Due to the immature status of the \mbox{driver/board} communication
mechanism and the hardware implementation, the bandwidth is limited at
2.3~GB/s, not showing the gain that should have hopefully resulted
from adoption of a \pcie Gen3 platform.

We are however confident to be able to improve on these results once
we employ more of the advancements that the new platform has put in
place.
For example, we mention that in this first version we are not using in
parallel the DMA engines that the PLDA QuickPCIe Expert provides; in
\apenetp, a similar optimization was introduced in a second stage of
development and led to a performance gain of about 40\% in
bandwidth~\cite{ammendola:2013:reconfig}.
Moreover, it is likely that current software implementation of the
driver will need significant optimization so as not to become the new
bottleneck in reaching peak performance: we are considering moving a
number of driver tasks from kernel to user space; for example,
offloading memory mapping \texttt{tx\_ring} and \texttt{event queue}
would reduce the switches between user and kernel spaces and help us
reach the Gen3 bandwidths.

\acknowledgments

This work was partially supported by EU Framework Programme 7 EURETILE
project, grant number 247846; Roberto Ammendola and Michele Martinelli
were supported by MIUR (Italy) through INFN SUMA project.

\bibliographystyle{JHEP}
\bibliography{../../ape_bib/bibliography}

\end{document}